\documentstyle[12pt,epsf,here]{article}
\input{epsf}                                                               
\begin{document}
\begin{center}

{\Large \bf Backward emitted high-energy neutrons in hard reactions of 
p and $\pi^{+}$  on carbon}
\end{center}
\vspace{0.2cm}

\begin{center}
{A. Malki$^a$, J. Alster$^a$, G. Asryan$^{c,b}$,
D. Barton$^c$, V. Baturin$^{e,d}$, N. Buchkojarova$^{c,d}$,
A. Carroll$^c$, 
A. Chtchetkovski$^{e,d}$,
S. Heppelmann$^e$,
 T. Kawabata$^f$,
A. Leksanov$^e$,
Y. Makdisi$^c$, E. Minina$^e$, I. Navon$^a$,
H. Nicholson$^g$,
Yu. Panebratsev$^{h}$,
E. Piasetzky$^a$, 
 S. Shimanskiy$^{h}$,
A. Tang$^i$, J.W. Watson$^i$, 
H. Yoshida$^f$, 
 D. Zhalov$^e$}
\end{center}
\vspace{0.2cm}

{\it $^a$School of Physics and Astronomy, Sackler Faculty of Exact
Sciences, Tel Aviv University.} 
{\it $^b$Yerevan Physics Institute, Yerevan, Armenia.}
{\it $^c$Brookhaven National  Laboratory.} 
{\it $^d$Dept. of Physics, St. Petersburg Univ., St. Petersburg, Russia.}
{\it $^e$Physics Department, Pennsylvania State University.} 
{\it $^f$Dept. of Physics, Kyoto Univ., Kyoto, Japan.}
{\it $^g$Mount Holyoke College.} 
{\it $^h$J.I.N.R., Dubna, Russia.}
{\it $^i$Dept. of Physics, Kent State University.}
\vspace{0.3cm}

\vspace{0.2cm}

{\bf Abstract}
\vspace{0.2cm}

Beams of protons and pions of 5.9 GeV/c  were incident on a C 
target.
Neutrons  emitted into the back hemisphere, in the 
laboratory system, were detected in (triple) coincidence with two emerging 
$p_t>$0.6 GeV/c particles.  We present the momentum spectra of the
backward going neutrons, which have the same 
 universal shape 
observed in earlier (inclusive)  reactions induced by 
hadrons, $\gamma$, $\nu$, and $\bar{\nu}$ beams.
We also integrated the spectra and determined the fraction of 
the hard scattering events which are in coincidence with at least
one neutron emitted into the back hemisphere, with momenta above 0.32
GeV/c. 
 Contrary to the earlier measurements which found that only a 
small fraction (of the order of 10$\%$) of the 
total inelastic cross section for light nuclei was associated with
backward going 
nucleons, we find that about half of the events are of this nature. 
We  speculate that the reason for the large difference
is the strong total center of mass (s) dependence of the hard reaction and 
short range nucleon correlations  in nuclei. 

\vspace{0.3cm}


\newpage

\vspace{1.0cm}

 There has been an intensive experimental program directed toward the
systematic characterization of the emission of backward going nucleons
from nuclei 
in  collisions with high energy ($>$ 1 GeV) hadrons
\cite{kn:I20,kn:I11,kn:I12,kn:I17}, 
real photons \cite{kn:I21}, 
virtual photons \cite{kn:I20}, neutrinos 
and antineutrinos \cite{kn:I4,kn:I5,kn:I6}. 
In these  experiments  the nucleons were emitted into angles larger than
90$^o$ in the laboratory system.
Thus, the kinematical conditions were
 such that the observed backward nucleon could not result from a
single two-body
scattering of the incident particle with a  nucleon at rest in the
nucleus.
Even though these experiments involved a large variety of interactions,
energies and nuclei,
a common universal spectrum of the backward nucleons
was observed. This spectrum can be parameterized 
by the expression: 
\begin{equation}
(E/p){d\sigma\over d(p^2)} = C e^{-Bp^2}
\label{energy}
\end{equation}
where $E$ and $p$ are the energy and momentum of the backward going 
nucleon. 

 For incident beams above about 1 GeV/c and for backward going 
nucleons  above about 0.3 GeV/c
the slope parameter $B$ was found to be
independent of incident energy and beam type and  target
nucleus and only weakly dependent on the angle of the backward
going particle ( see below for details).
The absolute scale parameter $C$ depends on the nucleus and only weakly on 
the incident energy and projectile.
The fraction of high energy (above
the Fermi sea level) backward nucleons  with
respect to the total inelastic events 
for light nuclei
(C, Ne) is of the order of 10$\%$ (see Table I).

   In this paper we present the first measurement of high energy backward
neutrons emitted from a nucleus {\it in coincidence  with two high-$p_t$
particles}, ($p_t >$ 0.6 GeV/c). 
We find that, while the universal shape of the
momentum spectra 
is maintained, the measured fraction of
events with two high $p_t$ particles in coincidence with a
backward going
neutron is substantially higher than the ratios measured for the more
inclusive reactions.
A detailed description of our
experiment follows.

We   present the first results from a measurement which  
was performed during 1998 with the rebuilt EVA spectrometer at the AGS
accelerator
of Brookhaven National 
Laboratory. 
The spectrometer consisted of a super-conducting solenoidal
magnet operated 
at 0.8 Tesla (see Fig. 1).   The scattered particles were tracked by 4
sets of
4-layer straw tube cylindrical  drift 
chambers (not shown) which surrounded the beam axis cylindrically. These
straw tubes measured  
 the transverse momenta of the
particles by drift time and the polar scattering angle by charge division.
Details on EVA spectrometer straw tube system are given in 
refs. \cite{kn:ref3,kn:ref4,kn:ref5,kn:ref7}. The major
change to the spectrometer in addition to the improved performance of the
solenoid and straw tubes was the installation of two new neutron counter
arrays which increased the acceptance by a factor of 2.5  over the
1994 configuration \cite{kn:haim}.

At a momentum of 5.9 GeV/c the beam consisted of about 40$\%$ 
protons and 60$\%$ pions, identified by  
a sequence of two differential
Cerenkov counters.
The beam entered along the symmetry axis ($z$) of the magnet with an
intensity of $\approx$1$\times 10^7$  particles over a one second
spill,
every 3 seconds.
A scintillator hodoscope
in the beam served as timing reference.  
Three  solid targets consisting of various combinations of CH$_2$ and C
were placed on the $z$ axis,
separated
by about 20 cm. They were 5.1$\times$5.1 cm$^2$ wide and 6.6 cm long
in the
$z$ direction.  Their positions were interchanged several times at regular
intervals. Some of the runs were with three  C targets and some with 
two C targets and one CH$_2$ target.

As indicated in Fig 1, we triggered the spectrometer on two positively
charged particles which emerged from the downstream end of the solenoid
 at polar angles of (27.5$\pm$3)$^o$ which corresponds to about 90$^o$
in the pp center of mass.
For this analysis, events with two particles with a $p_t >$ 0.6 GeV/c
which 
originated from one of the C targets were selected.  The trigger required
that one particle go to the left of the beam, and the other to the right.
In addition we required that there were no additional charged tracks in
the
straw tubes.  The polar angle coverage of the inner straw tube cyclinder
extended from
about 10$^o$  to 150$^o$.
Three scintillator arrays
measured 
the direction and energy of neutrons,
in coincidence with these two particles.

In Fig. 1 we present a schematic picture of the setup. We show  the 
magnet of the EVA spectrometer and the positions of the targets.
Below the targets we placed a series of 12 scintillator bars   (ARRAY 1 
in Fig. 1) covering an area of 0.6$\times$1.0 m$^2$ and 0.25 m (2 layers
$0.125$ m each)
deep.  They spanned a polar angular range of 84$^o$ to 110$^o$
and an azimuthal range from 165$^o$ to  195$^o$.
A similar array of 16 scintillator bars (ARRAY 2), covering an area 
of 0.8$\times$1.0 m$^2$ and 0.25 m deep,
spanned a polar angular range of 
110$^o$ to 132$^o$  and the same azimuthal range as the first one. 
Each individual counter 
in these two arrays was  10$\times$12.5$\times$100 cm$^3$ in size
and 
had a 5.1 cm photomultiplier at each end.
The third array (ARRAY 3) 
was constructed from one layer of eight   10$\times$25$\times$100  cm$^3$
counters with a
12.7 cm photomultiplier at each end.
This third array  covered an area of 2.0$\times$1.0 m$^2$
and spanned a polar angular range of 72$^o$ to 120$^o$
and an azimuthal range from 120$^o$ to  150$^o$.  
A set of veto
counters (not shown  in Fig 1.) served to discriminate against charged
particles. 
Lead sheets (not shown  in Fig 1.) with a total thickness of 1.7 
radiation lengths were placed in front of the veto counters in order to
reduce the number of photons entering the time of flight (TOF) spectrum.
All counters
were 
set to an electron equivalent detection threshold of 1 MeV by fixing a
discriminator at the Compton edge of a $^{60}$Co gamma source. This 
 procedure is similar to the description  in 
 ref. \cite{kn:addref1}. 
The detection efficiency was determined by the Monte
Carlo method
described in ref \cite{kn:j1}. The efficiencies depend on the neutron
momentum and they ranged from about 30$\%$ at 0.15 GeV/c to about 15$\%$
at 0.5 GeV/c, for a typical single counter. We  considered only neutrons
above 0.15 GeV/c to avoid large   
uncertainties in the efficiency calculations of the neutrons at low
momenta. A fraction of the neutrons gets absorbed
 on the trajectory from the
target
to the counters.  The probability that a neutron was removed was
calculated by assuming that
the removal cross section was equal to the non-elastic cross sections in
the  materials \cite{kn:addref2}.
We assigned an uncertainty
of 25$\%$   to these removal cross sections. The attenuation  depends
on the neutron momentum and the values ranged from about 35$\%$ at the
low momenta to about 20$\%$ at the high momenta.
The  neutron momenta were determined from their TOF.
  A clearly
identified peak due to remaining
photons from the targets was used for calibration and to measure the
timing resolution. That resolution was
$\sigma \le$ 1 ns which corresponds
to a momentum
resolution of $\sigma$= 30 MeV/c at the highest momentum (0.5 GeV/c).    
 We applied a cut  in the TOF spectrum at 7 ns/m, keeping neutrons 
below 0.5 GeV/c, in order to eliminate the remaining photons. 
 
 In Fig 2. we present the measured invariant momentum spectra 
$(E_n/p_n)\times{N_3\over d(p_n^2)}$ in arbitrary units 
for pion and proton incident beams, where
$E_n$ and $p_n$ are the energy and momentum of the neutron detected in 
the backward hemisphere 
($90^o<\theta_n<130^o$) and $N_3$ is defined below. We call  
$N_2$  the number of events with  two high  transverse momentum 
charged
particles with $p_t>0.6$ GeV/c, each and no other charged particles
seen in the detector.
 We 
applied software cuts to allow better determination of the target position
from the track reconstruction 
and to obtain a better separation between incident protons and
pions.   The  number of triple coincidence events which fulfill all the conditions of 
the $N_2$ events and, in addition,  have a single neutron in the
scintillator bars  is indicated by $N_3$.  
Since the efficiency and attenuation corrections depend on the
neutron 
momentum they were done event by event. The resulting  spectra are 
plotted  on a semi-logarithmic scale as a function of ${p_n}^2$. 
The error bars represent the  
statistical errors only. The curves  above ${p_n}^2>0.1$ (GeV/c)$^2$ 
were fitted to a first 
order 
polynomial. The slope parameters ($B$ in equation 1) for the proton
and pion incident beams are shown in the figure with their fitting error.
For comparison we quote the slope parameter obtained from the
p+C$\rightarrow$ n+X
data \cite{kn:I12} at 7.5 GeV/c. Their value at 119$^o$, which lies
within the range of our measurement, is B=13.0$\pm$0.8 (GeV/c)$^{-2}$. 
The neutrino measurements \cite{kn:I4,kn:I5,kn:I6} 
report a value of $B$=9.5-10.7 (GeV/c)$^{-2}$ for protons emitted into the
whole backward
hemisphere 
 with uncertainties  ranging from 0.3 to 2 (GeV/c)$^{-2}$.
Ref. {\cite{kn:I4} presents a compilation of other experiments at 
several different  energies of  hadron and photon beams incident on
a variety of targets, including C and other light nuclei. For 
particles in the 
common angular range of
120$^o$-150$^o$ of the different experiments, 
the slope parameters all lie within the range of  10-12.5 (GeV/c)$^{-2}$
with a typical
error of 2 (GeV/c)$^{-2}$. Our first conclusion is that the slopes we
measured in this experiment agree, within the measured uncertainties, with 
the slopes obtained using hadrons, 
photons and neutrino beams of different energies incident on various targets. 
The angular ranges are not the same for all experiments but this does not
modify the values of $B$ sufficiently to affect the conclusion 
(see \cite{kn:I6}).

We also wish to compare our yield of  backward
scattered neutrons above 0.32 GeV/c to those of the other experiments.
In Fig. 3 we show the (triple coincidence) backward yield per unit solid
angle divided by 
the (double coincidence) two high-$p_t$ particle yield vs. the
neutron angle.  Each point in this figure contains 10-20 
combinations of  target positions and a neutron-counter
(with the exception of the most forward angle point which includes 3 
such combinations only). The errors in the figure 
are the statistical errors combined with up to 10$\%$ systematic
errors due to software cuts, 
uncertainties  in the neutron detection thresholds and
uncertainties related to determining the exact geometrical positions of
the
counters.
To obtain the yield into the backward hemisphere 
we fitted the ratios  to a
constant (see fig 3). 
For the  proton induced reaction this gives a value of 
(7.4$\pm$0.4)$\%$ /sr  and for the   pion induced reaction the value of 
(6.5$\pm$0.6)$\%$/sr   above 
$90^o$ and $100^o$,
 respectively where the ratios become constant.
We used the parameters of the fit to extrapolate up to $\theta_n=180^o$,
assuming that the ratio remains constant. 
The results for the pion and proton induced reactions are shown in Table 1
compared with the  earlier mentioned experiments. Our results are
given for the integration up to $\theta_n=130^o$ with the proper
error as well as for the integration to $\theta_n=180^o$ which assumes
the constant ratio out to that angle. 
Our second conclusion from this experiment is, that  in
this measurement the ratio is substantially larger (3-4 times) 
than for the more inclusive measurements. This is true even if we 
include only our yield out to  $\theta_n=130^o$. Integrating out to
180$^o$ can only increase that ratio.

The earlier experiments  have led to several 
theoretical interpretations. The models that have been  discussed
can be divided into two  main classes. The first class  
\cite{kn:I14,kn:I15,kn:I18,kn:I19} deduces from 
the universality that the measurement must provide direct information on
the 
nuclear ground state, especially on the high-momentum part of the wave 
function. These models 
assume that the 
projectile interacts with a single nucleon and that the backward yield 
of nucleons is due to the  
break-up of pre-existing two, or more, 
correlated nucleon 
clusters \cite{kn:I14,kn:I15}. One of the more extreme models  in
this class assumes that the high 
momentum of the 
struck nucleon is balanced coherently by the residual nucleus
\cite{kn:I18,kn:I19}.
A second class of models \cite{kn:I16,kn:I17} assumes that the backward 
yield is due to rescattering processes in the nucleus of the incident and
outgoing 
particles. These initial and final interactions include 
true $\pi$-absorption and $\Delta$-rescattering in the intermediate
states.

The results of our experiment that the ratio of backward going neutrons
above a momentum of 0.32 GeV/c in coincidence with two high p$_t$ protons,
is substantially larger adds new information that has to be accomodated in
the theoretical models. 
We speculate that the reason for the difference is the strong total center
of mass (s) dependence of the hard reaction cross section and it's
sensitivity to the short range nucleon correlations in nuclei. The strong
s dependence of the hard reaction (for example 1/s$^{10}$ for pp elastic
scattering) selectively chooses the high momentum 
 protons  in the nuclei.
Those protons most likely have a correlated partner at short range which
are the neutrons that dominate the backward going yield
\cite{kn:I14,kn:I15,kn:haim}.  This speculation needs to be checked with
detailed calculations.

We wish to thank Drs. L. Frankfurt M. Strikman and M. Sargsyan for their 
theoretical input. 
We are pleased to acknowledge the assistance of 
the AGS staff in building and rebuilding the detector and supporting 
the experiment, particularly our liaison
engineers, J. Mills, D. Dayton, C. Pearson. We acknowledge the continuing
support of  D. Lowenstein and P. Pile. 
This research was supported by the U.S. - Israel Binational Science
foundation, the Israel Science Foundation founded by the Israel Academy of
Sciences and Humanities, the NSF grants PHY-9501114, PHY-9722519 and
the U.S. Department
of Energy grant DEFG0290ER40553.\\\\

\newpage
{\bf Tables}
\begin{center}
\textbf{Table 1}
\vspace*{1 cm}
\end{center}
\begin{tabular}{|c|c|c|c|c|c|c|} \hline
incident/ &energy&target&integration&integration&Ref.&RATIO\\
backward emitted&&&range&range&&\\
particle&(GeV)&&(GeV/c)&(Deg.)&&($\%$) \\ \hline
p/n&5.9&C&0.32-0.5&90-130&a &29.9$\pm$1.6\\ \hline
$\pi$/n&5.9&C&0.32-0.5&90-130&b&26.2$\pm$2.4\\ \hline
$\bar{\nu}$/p&$WB^c$&Ne&0.2-0.7&90-180&d1&7$\pm$1\\ \hline
p/p&3.66&Ne&0.2-0.7&90-180&d2&8$\pm$3\\ \hline
$\bar{\nu}$/p&$WB^e$&Ne&0.2-0.8&90-180&f&10$\pm$1\\ 
$\nu$/p&&&&&&12$\pm$4\\
NC/p&&&&&&10$\pm$1\\ \hline
$\bar{\nu}$/p&$WB^e$&Ne&0.35-0.8&90-180&g&6.0$\pm$0.2\\
$\nu$/p&&&&&&8.5$\pm$0.3\\ \hline
p/p&0.64&C&0.311-0.54&90-180&h&6.0$\pm$1.5\\ \hline
\end{tabular}
\vspace*{1 cm}\\
a- This work. If we extrapolate to 180$^o$ under the assumptions in the
text, we get a value of (46.5$\pm$2.5)$\%$.\\
b- This work. If we extrapolate to 180$^o$ under the assumptions in the    
text, we get a value of (40.8$\pm$3.7)$\%$.\\ 
c- A wide band $\bar{\nu}$ beam from 300 GeV/c incident protons. \\
d1- Results of ref. \cite{kn:I4}. \\
d2- Deduced in ref. \cite{kn:I4} from  measurements of ref. \cite{kn:I11}
to satisfy the same selection criteria as used in ref. \cite{kn:I4}.\\
e- A wide band $\bar{\nu}$ beam from 400 GeV/c incident protons. \\
f- "(BP+2BP)/TOTAL" from Table I of ref. \cite{kn:I5}\\
g- "1 Backward proton rate" from Table 1 of ref. \cite{kn:I6}.\\
h- Integral of I($\theta_3$) from Table 2 divided by $\sigma_t$
   from ref. \cite{kn:I17}

\newpage
{\bf Figure captions.}

\vspace{0.5cm}

{\bf Fig. 1} The  set up of the neutron counter arrays. 
Only the magnet
of the spectrometer
is shown with the position of the targets.
The lead shield 
and veto counters above the  neutron counter arrays are not shown.
\vspace{0.5cm}

{\bf Fig. 2} Proton and pion induced neutron invariant momentum spectra.
The vertical
axis is $ln[(E_n/p_n)\times{N_3\over d(p_n^2)}]$, The horizontal  
axis is ${p_n}^2$. $E_n$ and $p_n$ are the energy and momentum of the 
neutron. $N_2$ is the number of events with exactly two charged particles,
each with $p_t>0.6$ GeV/c, detected in the spectrometer. $N_3$ is the 
number of $N_2$ events that also have a single neutron entering the 
neutron  counters. The neutron yield is corrected for the detection
efficiency and attenuation. Above 
${p_n}^2>0.1$ (GeV/c)$^2$
the points are fitted to a straight line to obtain the 
slope parameter defined in Eq. 1. The resulting slopes  with the 
fitting errors are shown.
\vspace{0.5cm}

{\bf Fig. 3}  The relative yield per solid angle $dN_3\over N_2 d\Omega_n $
of backward going neutrons above 0.32 GeV/c as a function of the  
neutron angle. $N_3$ and $N_2$ are defined in Fig. 2 and the text.
The data are for the  proton and pion induced reactions. The lines 
represent fits to a constant which is used to estimate the total backward 
emission yield, see text.
\newpage

\begin{figure}[htbp]
\centerline{\epsfxsize=11.cm \epsfbox{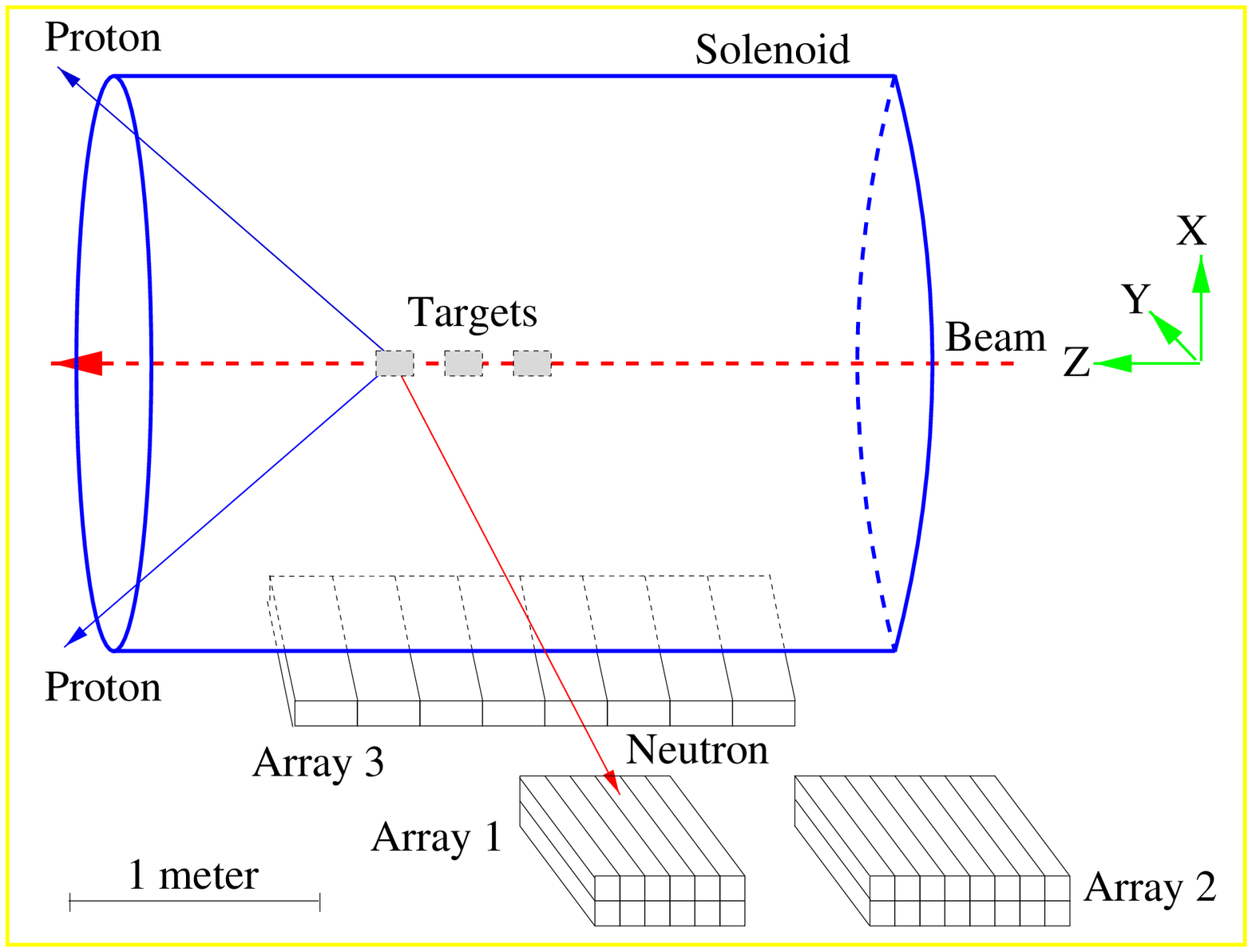}}
\vspace{1.0cm}
{FIG 1}
\end{figure}

\begin{figure}[H]
\centerline{\epsfxsize=11.cm \epsfbox{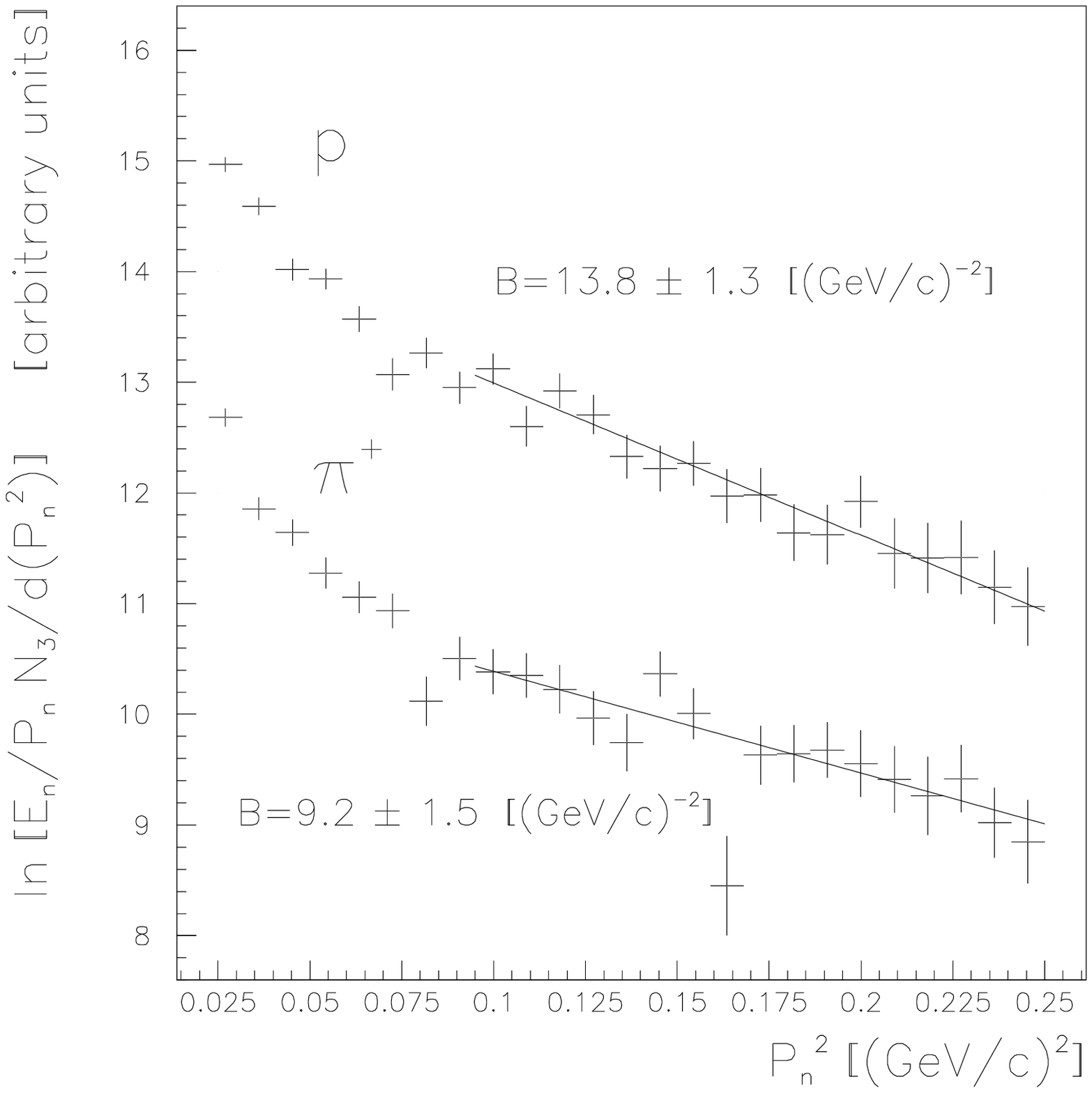}}
\vspace{0.5cm}
{FIG 2}
\end{figure}

\begin{figure}[H]
\centerline{\epsfxsize=11.cm \epsfbox{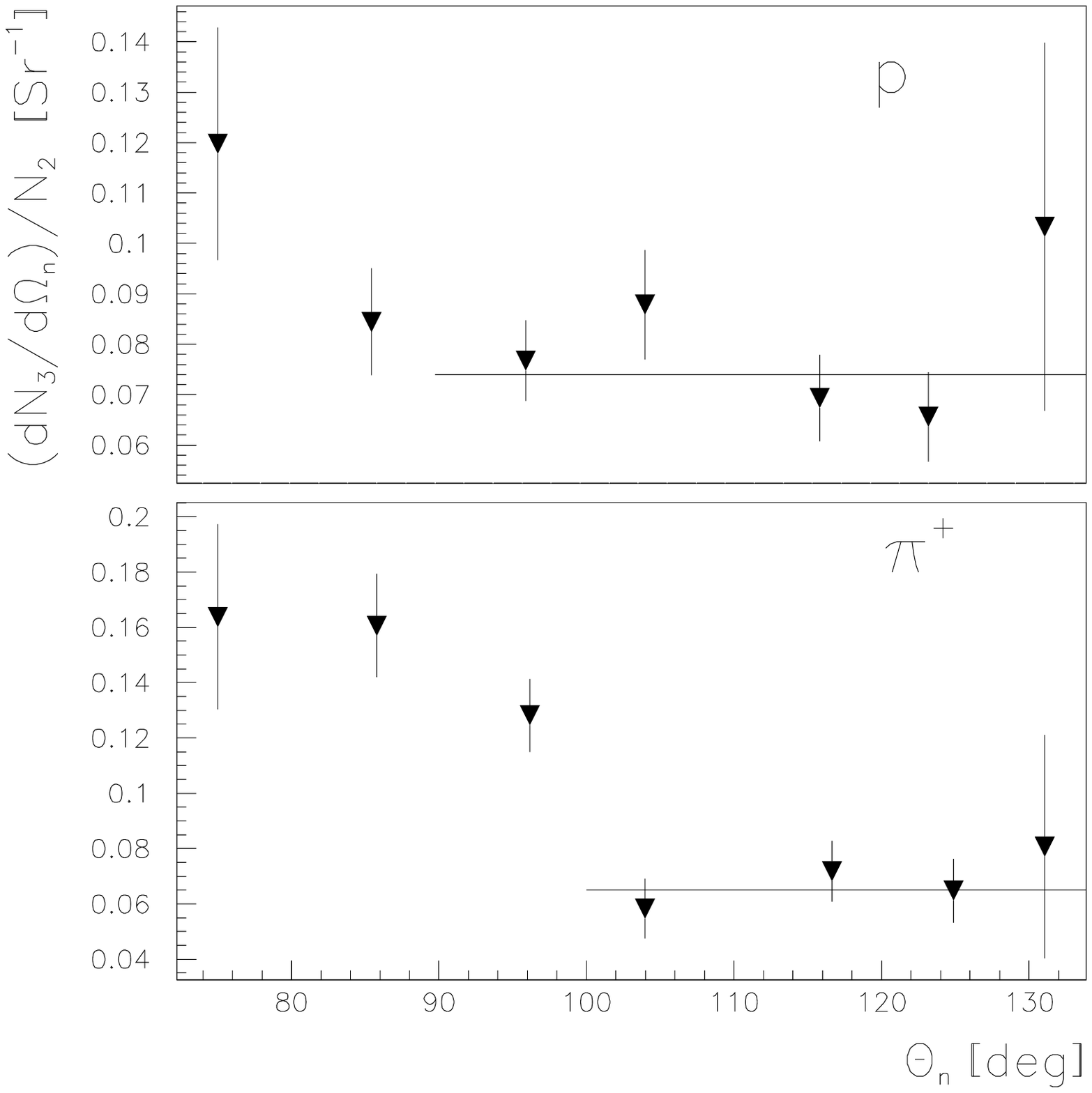}}
\vspace{0.5cm}
{FIG 3}
\end{figure}

\end{document}